\documentstyle[epsf]{elsart}
\begin{document}

\def\spose#1{\hbox to 0pt{#1\hss}}
\def\ltapprox{\mathrel{\spose{\lower 3pt\hbox{$\mathchar"218$}}
 \raise 2.0pt\hbox{$\mathchar"13C$}}}
\def\gtapprox{\mathrel{\spose{\lower 3pt\hbox{$\mathchar"218$}}
 \raise 2.0pt\hbox{$\mathchar"13E$}}}
\def\inapprox{\mathrel{\spose{\lower 3pt\hbox{$\mathchar"218$}}
 \raise 2.0pt\hbox{$\mathchar"232$}}}

\begin{frontmatter}

\begin{flushright}
{\normalsize FSU-SCRI-99-24}\\
{\normalsize hep-lat/9905028}\\
\end{flushright}

\title{Chiral Fermions on the Lattice}
\author{
Robert G. Edwards, Urs M. Heller and Rajamani Narayanan}
\address{
SCRI, The Florida State University, 
Tallahassee, FL 32306-4130, USA}

\begin{abstract}
The Overlap-Dirac operator provides a lattice regularization of massless
vector gauge theories with an exact chiral symmetry. Practical
implementations of this operator and recent results in quenched
QCD using this Overlap-Dirac operator are reviewed.
\end{abstract}
\end{frontmatter}

\section{Introduction}

Since the beginning of lattice gauge theory the regularization of chiral
fermions has been afflicted with severe problems. When regulating fermions
on a lattice, typically, unwanted doublers with opposite chirality appear.
These doublers can be lifted (given mass of the order of the cut-off)
at the cost of explicitly breaking chiral symmetry, as in the Wilson
fermion formulation. Alternatively, a remnant of chiral symmetry can be
retained, with a smaller number of doublers interpreted as flavors,
as in the staggered fermion formalism. However, at finite lattice spacing
the flavor symmetry is broken. Both these approaches fail from the
outset for regulating Weyl fermions. The central problem in the
non-perturbative regularization of gauge theories with Weyl fermions
is to write down a formula for the fermionic determinant when the
fermion is in some complex representation of the gauge group, since
depending on the topology of the gauge field the chiral Dirac operator
can be square or rectangular, where the difference between rows and
columns is the index.

Significant progress in the formulation of chiral gauge theories
has been made by the overlap formalism~\cite{overlap}.
The overlap formalism was inspired by two papers~\cite{inspire}. The central
idea is that an infinite number of Dirac fermions
(labeled by $s$) with a mass term of the form
$\bar\psi_s (P_L M_{ss^\prime} + P_R M^\dagger_{ss^\prime}) \psi_{s^\prime}$ 
and chiral projectors $P_{L,R}$ can be used to regulate a single Weyl
fermion if the infinite dimensional mass matrix $M$ has a single zero
mode but $M^\dagger$ has no zero modes. Kaplan's paper~\cite{inspire}
uses this idea to put chiral fermions on the lattice where they are
referred to as domain wall fermions since Kaplan used a mass matrix
that has a domain wall like structure.
In the overlap formalism the infinite number of Dirac fermions is described
by two non-interacting many body Hamiltonians, one for each side
of the domain wall, and the chiral determinant
is written as the overlap between their groundstates
\begin{equation}
\det {\mathrm C}(U) \Leftrightarrow  \langle 0- | 0+ \rangle .
\label{eq:detC}
\end{equation}
$|0- \rangle$ is the many body ground state of ${\mathcal H}^- =
a^\dagger \gamma_5 a$ and $|0+ \rangle$ the many body ground state of
${\mathcal H}^+ = a^\dagger H_w(U) a$, with $\gamma_5 H_w(U) = D_w(U)$
the usual Wilson-Dirac operator on the lattice with a fermion mass in
the supercritical region ($m_c < m < 2$). $a$ ($a^\dagger$) are canonical
fermion annihilation (creation) operators. On a finite lattice, the
single particle Hamiltonians are finite matrices of size $2K\times 2K$
with $K=V\times N \times S$ where $V$ is the volume of the lattice, $N$
is the size of the particular representation of the gauge group and $S$
is the number of components of a Weyl spinor. Then $|0-\rangle$ is made up
of $K$ particles. If $|0+\rangle$ is also made up of $K$ particles, then
the overlap is not zero in the generic case. If the background gauge field
is such that there are only $K-Q$ negative energy states for $H_w(U)$ then
the overlap is zero. Any small perturbation of the gauge field will not
alter this situation. Furthermore, the overlap
$\langle 0-| a^\dagger_{i_1}\cdots a^\dagger_{i_Q} |0+ \rangle$ will not
be zero in the generic case if the fermion is in the fundamental
representation of the gauge group showing that there is a violation of
fermion number by $Q$ units. So, clearly, the overlap definition of
the chiral determinant (\ref{eq:detC}) has the desired properties.

A generic problem with simulations of chiral gauge theories is that the
chiral fermion determinant is complex. A ``brute force approach'' is
feasible in the simulation of two dimensional models, and in this way the
overlap formalism has successfully reproduced non-trivial results in two
dimensional chiral models on the lattice~\cite{twod}. The brute force
approach, however, is clearly not feasible in four dimensions, where
efficient numerical techniques are essential. This prevented simulations
of chiral gauge theories and tests of the overlap formalism in four
dimensions so far.

Clearly, any formulation of lattice chiral gauge theories is also a
formulation of massless vector gauge theories with an exact chiral
symmetry and a positive fermion determinant (the product of the chiral
determinant for the left handed fermions and its complex conjugate for
the right handed ones). Lattice QCD using the overlap
formalism reproduces the well-known mass inequalities between mesons
and baryons, and the $U(N_f)_V\times U(N_f)_A$ symmetry in an $N_f$ flavor
theory is broken down to $U(1)_V\times SU(N_f)_V\times SU(N_f)_A$ by
gauge fields that carry topological charge~\cite{review}\footnote{See
section 9 of~\cite{overlap} for details.}. If the $SU(N_f)_V\times
SU(N_f)_A$ symmetry is spontaneously broken, then massless Goldstone
bosons should naturally emerge in the overlap formalism. Since the
symmetry breaking pattern is exactly as in the continuum, all the soft
pion theorems should hold on the lattice as well.

For a vector gauge theory the computation of the fermionic determinant
can be simplified significantly compared to the original version
(\ref{eq:detC}) that involves the computation of the overlap of two many
body ground states. One way to derive the simplified expression is to
start with the variant of domain wall fermions of ref.~\cite{Shamir},
%applicable for a vector gauge theory. Integrating out the heavy
applicable for a vector gauge theory. We choose this approach here to
emphasize the close connection between domain wall and overlap fermions.
%Integrating out the heavy
%fermion and the Pauli-Villars fields, Neuberger derived the following
%expression for the fermion determinant from the integration of the
%light fermions~\cite{herbert3}
Integrating out all the fermion and Pauli-Villars fields, Neuberger derived
the following expression for the determinant describing a single light
Dirac fermion~\cite{herbert3}
\begin{equation}
\det D_{DW}(\mu;L_s)=\det \left\{ {1\over 2} \left[1+\mu + (1-\mu) \gamma_5
 \tanh \left( -{L_s\over 2} \ln T_w \right) \right] \right\} .
\label{eq:detDW_Ls}
\end{equation}
Here $T_w$ is the transfer matrix in the extra direction, whose extent,
$L_s$, has been kept finite, and $0\le\mu\le 1$ describes fermions with
positive mass all the way from zero to infinity. In \cite{Shamir} the
fermion mass $\mu$ is denoted by $m_f$. In the limit
$L_s \to \infty$ (\ref{eq:detDW_Ls}) becomes
\begin{equation}
\det D_{DW}(\mu)=\det \left\{ {1\over 2} \left[1+\mu + (1-\mu) \gamma_5
\epsilon(-\ln T_w) \right] \right\} .
\label{eq:detDW}
\end{equation}
It is only in this limit that massless domain wall fermions have an exact
chiral symmetry.
Finally, taking the lattice spacing, $a_s$, in the extra direction to zero
one obtains the Overlap-Dirac operator of Neuberger~\cite{herbert1}
\begin{equation}
D(\mu)={1\over 2} \left[1+\mu + (1-\mu) \gamma_5\epsilon(H_w)  \right] .
\label{eq:Dmu}
\end{equation}

The external fermion propagator is given by 
\begin{equation}
{\tilde D}^{-1}(\mu)=(1-\mu)^{-1}\left[D^{-1}(\mu) -1\right]~.
\label{eq:prop}
\end{equation}
The subtraction at $\mu=0$ is evident from the original overlap
formalism~\cite{overlap} and the massless propagator anti-commutes with
$\gamma_5$~\cite{review,herbert1}. With our choice of subtraction
and overall normalization the propagator satisfies the relation
\begin{equation}
\mu \langle b^\dagger | \Bigl[ \gamma_5{\tilde D}^{-1}(\mu) \Bigr]^2
| b \rangle
= \langle b^\dagger | {\tilde D}^{-1}(\mu) | b \rangle
\ \ \ \ \forall \ \ b \ \ \ {\rm satisfying} \ \ \ 
\gamma_5 | b \rangle= \pm | b\rangle
\label{eq:Goldstone}
\end{equation}
for all values of $\mu$ in an arbitrary gauge field background~\cite{EHN1}.  
The fermion propagator on the lattice
is related to the continuum propagator for small momenta and small $\mu$ by
\begin{equation}
D_c^{-1} (m_q) = Z^{-1}_\psi {\tilde D}^{-1}(\mu)
\qquad {\rm with} \quad m_q  = Z_m^{-1} \mu
\end{equation}
where $Z_m$ and $Z_\psi$ are the mass and wavefunction renormalizations,
respectively. Requiring that (\ref{eq:Goldstone}) hold in the continuum
results in $Z_\psi Z_m =1$.
We find that a tree level tadpole improved estimate gives
\begin{equation}
Z_\psi = Z_m^{-1} = {2 \over u_0} \left[ m - 4 (1 - u_0) \right] ~,
\end{equation}
where $u_0$ is one's favorite choice for the tadpole link value. Most
consistently, for the above relation, it is obtained from $m_c$, the
critical mass of usual Wilson fermion spectroscopy. 

In the rest of this paper we discuss practical implementations of
the Overlap-Dirac operator and present some recent results. Owing to
the recent flurry~\cite{Niedermayer} of theoretical activity arising from the
``unearthing'' of the Ginsparg-Wilson relation~\cite{GW}, a few remarks are
in order. The massless Overlap-Dirac operator in (\ref{eq:Dmu}) satisfies
the Ginsparg-Wilson relation
\begin{equation}
D(0)\gamma_5 + \gamma_5 D(0) = 2 D(0)\gamma_5 D(0)
\label{eq:GW}
\end{equation}
implying that the massless propagator $(D^{-1}(0) -1)$ anticommutes with
$\gamma_5$. If we write 
\begin{equation}
D(0)={1\over 2} \left[1 + \gamma_5\hat H_a\right].
\label{eq:D0g}
\end{equation}
then the Ginsparg-Wilson relation reduces to $\hat H_a^2 = 1$. Since
we would want $\gamma_5 D(0)$ to be Hermitian, $\hat H_a$ should be a
Hermitian operator. With this reduction of the Ginsparg-Wilson relation, it
is easy to show that~\cite{GWover} 
\begin{equation}
\det D(0) = \left | \langle 0 - | 0 + \rangle \right |^2
\label{eq:overlap}
\end{equation}
{\it i.e.,} the overlap formula for a vector theory. Here $|0+\rangle$
is the many body ground state of $a^\dagger \hat H_a a$.
This establishes a one-to-one correspondence between the overlap
formula and the determinant of a fermionic operator satisfying the
Ginsparg-Wilson relation for massless vector gauge theories.

%Clearly, the choice for $\hat H_a$ is not unique. We choose
%$\hat H_a = \epsilon(H_w)$ as in Eq.~(\ref{eq:D0}) for
%the numerical results presented here. We note that $\gamma_5 H_w$ can
%be replaced by any improvement of the Wilson-Dirac operator.
%A different choice is obtained from the original domain-wall formalism.
%Starting from the variant of domain wall fermions of ref.~\cite{Shamir},
%applicable for a vector gauge theory, and integrating out the heavy
%fermion and the Pauli-Villars fields, Neuberger derived the following
%expression for the fermion determinant~\cite{herbert3}
%\begin{equation}
%\det D_{DW}(0;L_s)=\det \left\{ {1\over 2} \left[1 + \gamma_5
% \tanh \left( -{L_s\over 2} \ln T_w \right) \right] \right\} .
%\label{eq:detDW}
%\end{equation}
%Here $T_w$ is the transfer matrix in the extra direction, and the extent
%of the extra dimension, $L_s$, has been kept finite. In the limit
%$L_s \to \infty$ the $\tanh$ becomes an $\epsilon$ and comparing with
%(\ref{eq:D0g}) we find $\hat H_a = \epsilon(-\ln T_w)$. Since $\tanh^2 \ne 1$
%we see from (\ref{eq:detDW}) that domain wall fermions with a finite
%extra dimension do not satisfy the Ginsparg-Wilson relation and break
%the chiral symmetry. Numerically, domain wall fermions with a finite
%extra dimension are treated as $(d+1)$-dimensional fermions.

\section{Practical Implementations of $\epsilon(H_w)$}

In order to compute the action of $\epsilon(H_w)={H_w\over |H_w|}$ on
a vector one can proceed in several different ways. Since we are
interested in working in four dimensions, it is not practical to store
the whole matrix $H_w$. Therefore standard techniques to deal with the
square root of a matrix~\cite{brute} will not be discussed. 

One could attempt to solve the equation $\sqrt{H_w^2} \phi = H_w b$ to
obtain $\phi = \epsilon(H_w) b$ using iterative techniques.  Such
techniques have been developed to solve linear systems with fractional
powers of a positive definite operator using Gegenbauer
polynomials~\cite{Bunk} and applied to the Overlap-Dirac operator
in \cite{EHN2}.  

Another approach is to efficiently approximate $\epsilon(H_w)$ as a
sum of poles:
\begin{equation}
\epsilon (H_w) \approx g_N(H_w) = H_w \left [ c_0 + 
  \sum_{k=1}^N {c_k\over H_w^2 + d_k} \right ] ~.
\label{eq:epspole}
\end{equation}
The action of $\epsilon(H_w)$ on a vector then
involves a single conjugate gradient with multiple shifts~\cite{Jegerlehner}.

One approximation, called the polar decomposition~\cite{pensacola}, has been
adapted in this context and first used in the study of the three dimensional
Overlap-Dirac operator by Neuberger~\cite{herbert}. Here the
coefficient $c_0=0$ and
\begin{equation}
c_k = {1\over N\cos^2 {\pi\over 4N}(2k-1)};\qquad
d_k = \tan^2 {\pi\over 4N}(2k-1) ~.
\label{eq:polar}
\end{equation}
In this approximation,
\begin{equation}
\epsilon(z) \approx 
  g_N(z) = { (1+z)^{2N} - (1-z)^{2N} \over (1+z)^{2N} + (1-z)^{2N}} ~.
\label{eq:tanh}
\end{equation}
Clearly $g_N(z) = g_N(1/z)$ and $g_N(1) =1$.
The error $\epsilon(z)-g_N(z)$ is strictly positive
and monotonically decreases from $z=0$ to $z=1$. 

\begin{figure}
\epsfxsize=3in
\centerline{\epsffile{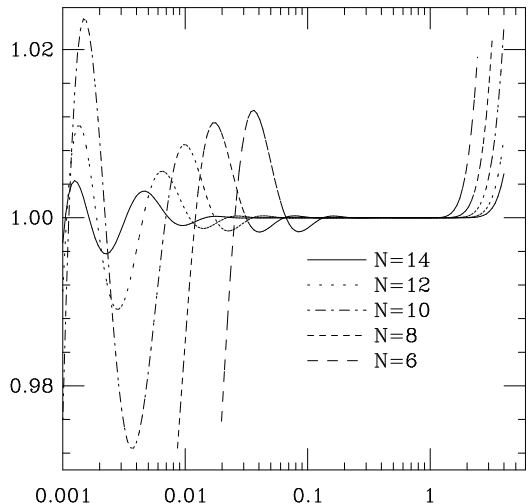}}
\caption{
Plots of the optimal rational function approximation to
$\epsilon(z)$ (Eq.~\ref{eq:epspole}) over the interval 
$[0.001,1]$ for $N=6$, $8$, $10$, $12$, and $14$. The small
$z$ region of $N=6$ and $8$ is not shown. The fits are still good for
some $z > 1$.}
\label{fig:approx}
\end{figure}

For another approximation, called the optimal
rational approximation~\cite{Remez}, the
coefficients are obtained numerically by an optimal fit using the Remez
algorithm. The coefficients in a slightly different notation have
been tabulated for $N=6,8,10$ in Ref.~\cite{EHN2}. We have found
it necessary to use $N=14$ for our recent applications.
The coefficients in this case are shown in Table~\ref{tab:coeff}.  
\begin{table}

\addtolength{\tabcolsep}{-1.0mm}

\begin{center}
\begin{tabular}{|cc|}
\hline 
(2.413330975e+00,1.361747338e+01) &
(6.257184884e-01,3.135687028e+00) \\
(2.925707925e-01,1.213113539e+00) &
(1.737405612e-01,5.596349298e-01) \\
(1.166359792e-01,2.752627333e-01) &
(8.372555094e-02,1.364115846e-01) \\
(6.216038074e-02,6.543005714e-02) &
(4.652496186e-02,2.923946484e-02) \\
(3.423610040e-02,1.164228894e-02) &
(2.404754621e-02,3.887745892e-03) \\
(1.545550091e-02,9.937321442e-04) &
(8.436481876e-03,1.684882417e-04) \\
(3.419245947e-03,1.585925699e-05) &
(1.138166539e-03,5.914114023e-07) \\
 \hline
\end{tabular}
\end{center}
\vspace{5mm}
\caption{The coefficient pairs $(c_k,d_k)$ $k=1,\dots,N$ for the $N=14$ optimal
rational approximation for which $c_{0} = 0.0850910$. The coefficients
are obtained as a result of an optimal fit over the interval $[0.001,1]$.
}
\label{tab:coeff}
\end{table}
In the
optimal rational approximation, the approximation to $\epsilon(z)$ has
oscillations and is not bounded by unity. A plot of the approximation
$g_N(z)$ obtained as a fit over the region $[0.001, 1]$ is shown in
Fig.~\ref{fig:approx} for $N=6$ to $14$. While we fit over this
region, the approximation is still good for $z$ somewhat larger than 1.
The approximation is bounded by unity only if $0.025 < z < 1.918$ for
$N=14$. In this range the maximum deviation from unity is equal to
$3.1\times 10^{-5}$.  This range will increase if one increases the
order of the approximation. For the current applications we found this
range to be sufficient.

The approximation to $\epsilon(H_w)$ by poles involves a multi-shift
inner conjugate gradient and therefore it seems necessary to store $N$
vectors where $N$ is the order of the approximation. One can avoid
storing the extra vectors if one is willing to perform two passes of
the inner conjugate gradient~\cite{twopass}.  A Lanczos based
algorithm that also avoids this extra storage by requiring two passes
has been proposed, but it involves an explicit diagonalization of a
tridiagonal matrix~\cite{Borici}. In this method $H_w$ is approximated
by a small dimensional tridiagonal matrix (anywhere between $100$ and
$1000$) and $\epsilon(H_w)$ is computed by first diagonalizing the
tridiagonal matrix and then performing the trivial operation of
$\epsilon$ on the eigenvalues.  The accuracy is increased by
increasing the order of the tridiagonal matrix.

Since in practice it is the action of $D(\mu)$ 
on a vector we need, we can check for the convergence of the complete
operator at each inner iteration of $\epsilon(H_w)$. This saves some
small amount of work at $\mu=0$ and more and more as $\mu$ increases,
while at $\mu=1$ (corresponding to infinitely heavy fermions,
c.f.~Eq.(\ref{eq:Dmu})) no work at all is required.

\begin{figure}
\epsfxsize=4in
%\centerline{\epsfbox[100 150 500 450]{rho0_scale_use.ps}}
\centerline{\epsfbox[85 150 485 450]{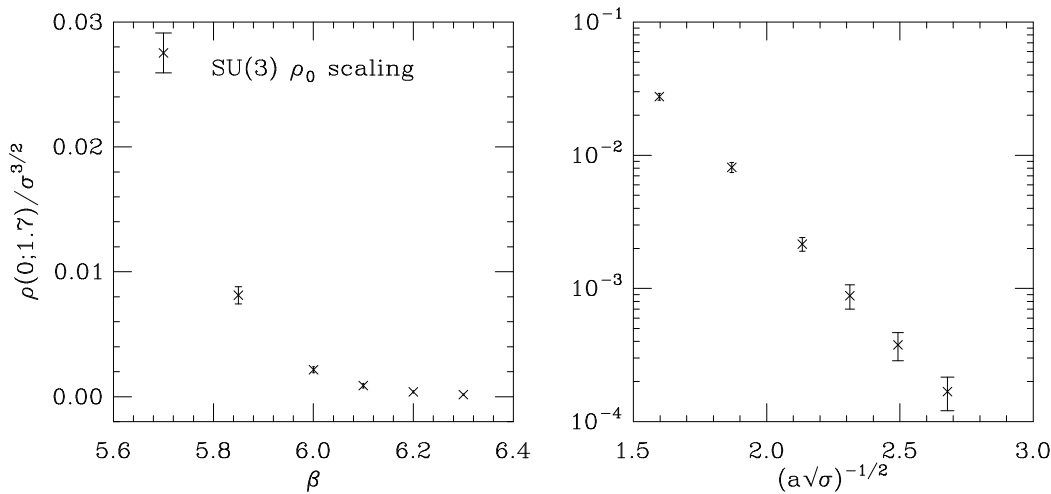}}
\caption{The approach of $\rho(0;1.7)$ to the continuum limit versus
$\beta$ and versus $1/\sqrt{a^2\sigma}$.}
\label{fig:rho0_scale_use}
\end{figure}

Each action of $\epsilon(H_w)$ involves several applications of
$H_w$ on a vector with the number depending on the condition number
of $H_w(m)$ in the supercritical mass region. In
Figure~\ref{fig:rho0_scale_use} we show the density of (near)
zero eigenvalues for $m=1.7$. We see that while
$\rho(0;1.7)$ decreases rapidly as $\beta$ increases, it does not appear
to go zero at a finite lattice spacing. The second part of
Figure~\ref{fig:rho0_scale_use} emphasizes that $\rho(0;1.7)$ decreases
exponentially in some power of $1/a$ (the power here is not well
determined, but reasonably fits $1/2$). This result implies that on
a specific gauge background, $H_w$ could have an arbitrarily
large condition number due to a few small eigenvalues. This can make
the computation of $\phi = \epsilon(H_w) b$ expensive for all methods
considered. In addition some care is needed when using the approximation
of $\epsilon(H_w)$ by a sum over poles.
Clearly $\epsilon(H_w)$ can be replaced by $\epsilon (sH_w)$ where
$s>0$ is an arbitrary scale factor. We should choose the scale so that
the maximum eigenvalue of $sH_w$ is not above the range where the
approximation is deemed good. Having so chosen a value for $s$, we
need to deal with the low lying eigenvalues that fall outside the
range of the approximation to $\epsilon(H_w)$. We do this by computing
a few low lying eigenvalues and eigenvectors of $H_w$, for which we
then know the contribution to $\epsilon(H_w)$ exactly, and
projecting them out before applying the approximation to the
orthogonal subspace for which the approximation is good.
The number of eigenvalues that have to be projected out
will depend on the lattice coupling, the lattice size and the lower
end of the range of the approximation.  It will roughly increase with
the volume at a fixed coupling making it difficult to go to large
lattice volumes at strong coupling. However, the number of eigenvalues
that have to be projected out will decrease as one goes to weaker
coupling even at a fixed physical volume. This is because the density
of eigenvalues of $H_w$ near zero goes to zero as one goes to the
continuum limit~\cite{rough}. The Ritz functional method~\cite{ritz}
can be used to efficiently compute the necessary low lying eigenvalues
and eigenvectors of $H_w$.

For the case of domain wall fermions at finite extent $L_s$ in the
fifth direction, the degree to which $\tanh(-L_s (\ln T_w(m))/2)$
approximates $\epsilon(-\ln T_w(m))$ is determined by $L_s$ and the
eigenvalues of $T_w(m)$ near $1$.  One can show
analytically~\cite{overlap} that in a fixed gauge background a unit
eigenvalue of $T_w(m)$ and a zero eigenvalue of $H_w(m)$ occur at the
same mass $m$. Also, the change of the corresponding eigenvalue in $m$
is the same for both $T_w(m)$ and $H_w(m)$. This implies
that the density of zero eigenvalues $\rho(0;m)$ is the same for both
$H_w(m)$ and $-\ln(T_w(m))$. The degree to which these zero
eigenvalues affect physical results is determined by the physical
observable, $L_s$, and the fermion mass $\mu$.
Studies of the $L_s$ dependence for various
quantities at non-zero fermion masses can be found in
Ref.~\cite{Columbia}. In particular, larger $L_s$ at fixed mass $\mu$
is needed for stronger coupling due to the increasing $\rho(0;m)$.

\section{Spontaneous chiral symmetry breaking}

\begin{figure}
\epsfxsize=4in
%\centerline{\epsfbox[100 150 500 450]{cfg17.ps}}
\centerline{\epsfbox[85 150 485 450]{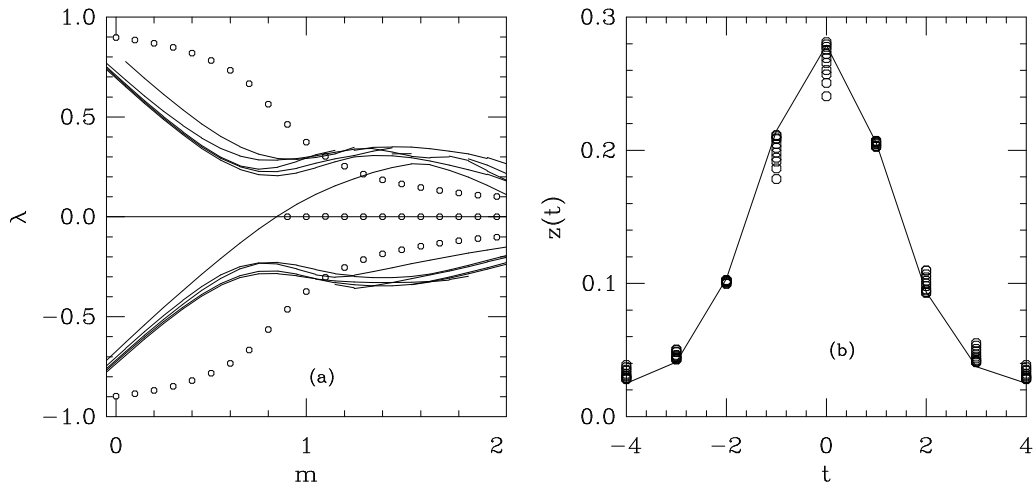}}
\caption{(a) The low lying spectrum of $H_o$ (octagons) in a pure
SU(2) gauge field background at $\beta=2.5$ together with the spectral
flow of $H_w$ (lines). The zero mode of $H_o$ found for $m > 0.8$
is singly degenerate and is associated with an instanton in the background
gauge field. (b) Shape of the zero mode of $H_o$ projected to one axis for
the various $m$ (octagons) and of $H_w$ at the crossing point (line).}
\label{fig:cfg17}
\end{figure}

Spontaneous chiral symmetry breaking is an important feature of QCD.
However, it is not fully realized on a finite lattice and at finite
quark mass. Thus one needs to carefully study the approach to the infinite
volume and chiral limit. Conventional lattice fermion formulations explicitly
break the chiral symmetry (at least partially) at finite lattice spacing,
obscuring the approach to the infinite volume and chiral limit in
practical simulations. Overlap fermions preserve the chiral symmetry at finite
lattice spacing. This should facilitate a study of spontaneous chiral
symmetry breaking. As a practical
test of the Overlap-Dirac operator, we consider quenched
QCD. The chiral limit of quenched QCD is tricky, though, because topologically
non-trivial gauge fields are not suppressed in this limit. 
Gauge field topology results in exact
zero modes of $D(0)$ as long as one is in the supercritical region of
$H_w(m)$. This is demonstrated in 
in Fig.~\ref{fig:cfg17} where we show the spectral flow of eigenvalues
of both $H_w(m)$ and $H_o = \gamma_5 D(0)$ as a function of $m$
for an SU(2) gauge background
at $\beta=2.5$ on an $8^4$ lattice. We see a single level crossing zero
near $m=0.9$ in the spectral flow of $H_w$. At this mass,
we see the sudden appearance of a zero eigenvalue (with chirality $1$)
among the smoothly changing non-zero eigenvalues (in opposite sign
pairs with chirality equal to their eigenvalue). 

Zero eigenvalues of $H_o$ due to global topology
have a definite chirality. The spectrum of $H_o$ is in [-1,1] and the
non-zero eigenvalues of $H_o$ that have a magnitude less than
one come in pairs,
$\pm\lambda$. The associated eigenvectors are not eigenvectors
of $\gamma_5$, but rather $\gamma_5$ has expectation value
$\pm\lambda$ in the eigenvectors, 
$\psi^\dagger \gamma_5 \psi = \pm\lambda$.
Since $H_o$ is an even dimensional matrix, the unpaired zero eigenvalues
have to be matched by unpaired eigenvalues equal to $\pm 1$. This is what
is expected to happen in a topologically non-trivial background.
It is straightforward to obtain the spectrum of $D(\mu)$ from the spectrum
of $H_o(0)$. Due to the continuum like spectrum one can
study the approach to the chiral limit using the Overlap-Dirac
operator by separating modes due to global topology from the remaining
non-zero eigenvalues~\cite{EHN1}.

The main quantity that needs to be
computed numerically is the fermion propagator $\tilde D^{ -1}(\mu)$
in Eqn.~(\ref{eq:prop}).  Certain properties of the Overlap-Dirac
operator enable us to compute the propagator for several fermion
masses at one time using the multiple Krylov space
solver~\cite{Jegerlehner} and also go directly to the massless limit.

We note that 
\begin{equation}
H_o^2(\mu) = D^\dagger(\mu) D(\mu) = D(\mu) D^\dagger(\mu) = 
\left(1-\mu^2 \right) \left[ H_o^2(0) + {\mu^2\over 1-\mu^2} \right]
\label{eq:H_o_mu}
\end{equation}
with
\begin{equation}
H_o^2(0) = {1\over 2} + {1\over 4} \left[\gamma_5\epsilon(H_w) +
 \epsilon(H_w)\gamma_5 \right]
\label{eq:H_o_0}
\end{equation}
Eq.~(\ref{eq:H_o_mu}) implies that we can solve the set of equations
$H_o^2(\mu) \eta(\mu) = b$ for several masses, $\mu$, simultaneously
(for the same right hand $b$) using
the multiple Krylov space solver described in
Ref.~\cite{Jegerlehner}. We will refer to this as the outer conjugate
gradient inversion.
It is easy to see that $[H_o^2(\mu),\gamma_5]=0$,
implying that one can work with the source $b$ and solutions
$\eta(\mu)$ restricted to one chiral sector. 

The numerically expensive part of the Overlap-Dirac operator is the
action of $H_o^2(0)$ on a vector since it involves the action of
$[\gamma_5\epsilon(H_w) + \epsilon(H_w)\gamma_5]$ on a vector. If the
vector $b$ is chiral ({\it i.e.} $\gamma_5 b = \pm b$) then, 
$[\gamma_5\epsilon(H_w) + \epsilon(H_w)\gamma_5] b  = [\gamma_5 \pm 1]
\epsilon(H_w)b$. Therefore we only need to compute the action of
$\epsilon(H_w)$ on a single vector.

To study the possible onset of spontaneous chiral symmetry breaking in
quenched QCD, we stochastically estimate, for a fixed gauge field
background,
\begin{equation}
{1\over V} \sum_{x} \langle \bar\psi(x) \psi(x) \rangle_A 
= {1\over V} {\rm Tr}[{\tilde D}^{-1}(\mu)] 
\label{eq:pbp}
\end{equation}
and average over gauge fields. We also compute stochastically
$\omega = \chi_\pi - \chi_{a_0}$
\begin{equation}
\omega =  
{2\over V} \langle {\rm Tr}(\gamma_5\tilde D)^{-2}(\mu) + 
{\rm Tr}\tilde D^{-2}(\mu)\rangle 
= 
{2\over\mu} \langle \bar\psi \psi \rangle
- 2 \langle {d\over d\mu} \langle \bar\psi \psi \rangle_A \rangle ~.
\label{eq:omega}
\end{equation}
For a derivation of the above equation we refer the reader to 
Ref.~\cite{EHN1}. 
Some simple manipulations yield the following relations
\begin{eqnarray}
\langle b | {\tilde D}^{-1}(\mu) | b \rangle  & = & {\mu\over 1-\mu^2} b^\dagger
\left ( \eta(\mu) - b \right ) \cr
\langle b | (\gamma_5\tilde D)^{-2}(\mu) + \tilde D^{-2}(\mu) | b \rangle 
& = & {2\mu^2\over (1-\mu^2)^2} \left(\eta^\dagger(\mu) - b^\dagger\right)
\left ( \eta(\mu) - b \right ) 
\label{eq:simple}
\end{eqnarray}
where
\begin{equation}
H_o^2(\mu)\eta(\mu) = b \qquad {\rm with} \qquad \gamma_5 b = \pm b \quad.
\end{equation}

As discussed in Ref.~\cite{EHN1}, it is appropriate to remove the
topological contributions to the above quantities in order to study
the onset of chiral symmetry breaking. For this
we first compute the low lying spectrum of 
$\gamma_5 D(0)$ using the Ritz functional method~\cite{ritz} . This gives us,
in particular, information about the number of zero modes and their 
chirality.
In gauge fields with zero modes we always find
that all $|Q|$ zero modes have the same chirality. We have not found
any accidental zero mode pairs with opposite chiralities.
We then perform a stochastic estimate in the chiral sector that
has no zero modes and double the result to get the total contribution to
$\langle \bar\psi \psi\rangle$ and $\omega$ excluding topology.
In this sector, the propagator 
is non-singular even for zero fermion mass. Given a Gaussian random
source $b$ with a definite chirality all we have to do
is solve the equation
$H^2_o(\mu) \eta(\mu) = b$ for several values of $\mu$. 

\begin{figure}
\epsfxsize=4in
%\centerline{\epsfbox[100 150 500 450]{omega_meson.ps}}
\centerline{\epsfbox[85 150 485 450]{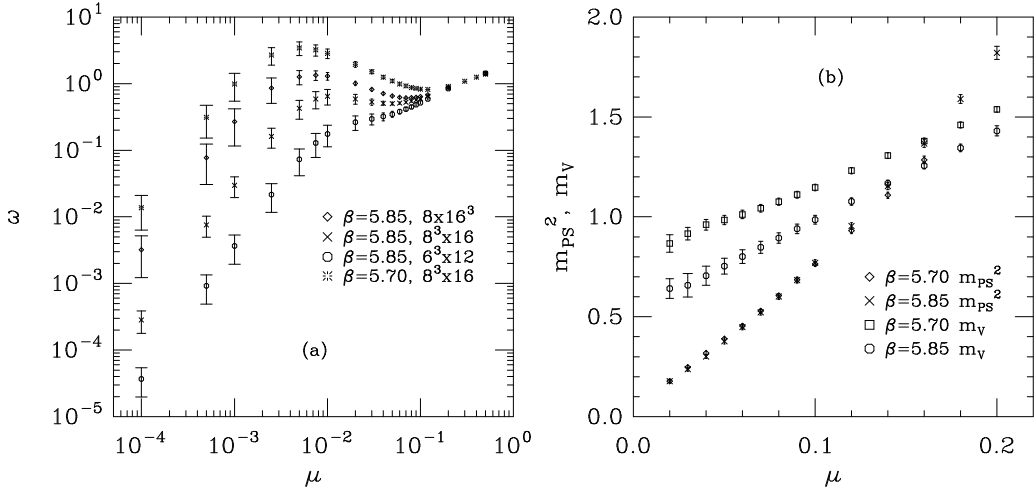}}
\caption{(a) $\omega$ without the contribution from global topology
as a function of $\mu$. (b) The pseudoscalar mass squared, $a m_{PS}^2$,
and the vector mass, $a m_V$, for $\beta=5.85$ and $5.70$ on an
$8^3\times 16$ lattice as a function of $\mu$.}
\label{fig:omega_meson}
\end{figure}

In Fig.~\ref{fig:omega_meson}a we show $\omega$ without the topology term
added for various lattice sizes and $\beta$ in SU(3) using 
$m = 1.65$. We see some indication of the onset of spontaneous chiral
symmetry breaking (with strong finite volume dependence) at
$\beta=5.85$ where, as $\mu$ decreases, there is a small region where
$\omega \sim 1/\mu$, then $\omega$ turns over and goes like
$\mu^2$. This latter behavior is expected in finite volume and is
obvious from the explicit $\mu^2$ dependence in Eq.~(\ref{eq:simple}).

We show in Fig.~\ref{fig:omega_meson}b pseudoscalar and vector meson masses
from a preliminary spectroscopy calculation for SU(3) $\beta=5.85$ and
$5.70$ on an $8^3\times 16$ lattice. Masses are extracted using
multiple correlation functions in an excited state fit.  The fermion
masses have been chosen to be above the region of decreasing $\omega$
from finite volume dependence in Fig.~\ref{fig:omega_meson}a, namely $\mu >
10^{-2}$.  As in the calculations for $\omega$ above, a multiple mass
shift conjugate gradient solver was used for several values of $\mu$
in the solution of $H^2_o(\mu) \eta(\mu) = b$ with chiral source $b$.
We see some slight deviation of $a m_{PS}^2$ from linearity for
decreasing $\mu$, and $a m_{PS}^2$ does not extrapolate to $0$ at 
$\mu = 0$ which we attribute to finite volume dependence. The vector
mass $m_V$ is fairly linear over the entire region.

\section {The Overlap-Dirac operator and random matrix theory}

\begin{figure}[t]
%\vspace*{-10mm} \hspace*{-0cm}
\begin{center}
\epsfxsize=4in
%\centerline{\epsfbox[100 150 500 450]{cfg17.ps}}
%\centerline{\epsfbox[85 150 485 450]{cfg17.ps}}
%\epsfxsize = 0.8\textwidth
\centerline{\epsfbox[100 175 550 500]{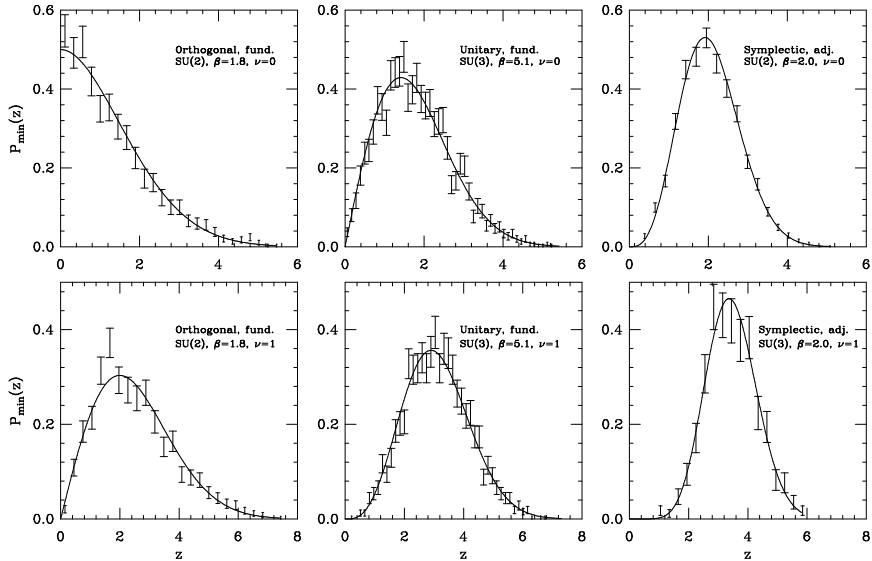}}
\end{center}
\caption{Plots of the probability distribution, $P_{\rm min}(z)$, of
the smallest (rescaled) eigenvalue $z$ for the orthogonal, unitary and
symplectic ensembles in the lowest two topological sectors. Here the
smallest eigenvalue $\lambda$ is rescaled by the volume $V$ and the
infinite volume chiral condensate $\Sigma$ to $z=\lambda V\Sigma$. The
curve in each plot is a fit to the prediction from random matrix
theory with the best value for the chiral condensate $\Sigma$.}
\label{fig:pmin}
\end{figure}

The Goldstone pions, associated with the spontaneous breaking of chiral
symmetry dominate the low-energy, finite-volume scaling behavior of the
Dirac operator spectrum in the microscopic regime, defined by
$1/\Lambda_{QCD} << L << 1/m_\pi$, with $L$ the linear extent of the system.
The properties in this regime are universal and can be characterized
by chiral random matrix theory (RMT) within three ensembles, depending
on some symmetry properties of the Dirac operator, and according to the
sector of fixed topology, entering via the number of exact zero modes
(see \cite{Verbaar} for a recent review).
Since the Overlap-Dirac operator has the same chiral properties as the
Dirac operator in the continuum, and since it has exact zero modes in
topologically non-trivial gauge fields, it is well suited to test the
predictions of RMT. In Figure~\ref{fig:pmin} the distribution of the
lowest (non-zero) eigenvalue is compared to the predictions of chiral RMT
for examples in all three universality classes -- SU(2) in the fundamental
representation for the orthogonal ensemble, SU(3) in the fundamental
representation for the unitary ensemble and SU(2) in the adjoint
representation for the symplectic case -- and in the sectors with zero or one
exact zero modes~\cite{ov_RMT}. Excellent agreement is seen. In addition,
the condensate $\Sigma$ obtained in the two different sectors of each
ensemble from fits to the RMT predictions agreed within errors. This agreement
further validates the chiral RMT predictions on the one hand and
strengthens the case for the usefulness of the overlap regularization of
massless fermions on the other hand.

\section {Conclusions}

The Overlap-Dirac operator provides a formulation of vector gauge theories
on the lattice with an exact chiral symmetry in the massless limit and
no fermion doubling problem. The use of the Overlap-Dirac operator is,
however, CPU time intensive. We reviewed a few methods
to implement the operator acting on a vector. Of these, we found the optimal
rational approximation method, in conjunction with the exact treatment
of a few low lying eigenvalues and eigenvectors of $H_w$ in
$\epsilon(H_w)$, the most efficient. Further improvements in the
numerical treatment of the Overlap-Dirac operator would be very helpful.

The Overlap-Dirac operator has exact zero modes with definite chirality
in the presence of topologically non-trivial gauge fields. Due to their
good chiral properties overlap fermions are well suited for the study
of spontaneous chiral symmetry breaking. It is possible to separate
the contribution of the exact zero modes due to topology in a numerical
computation. We have presented sample results in a quenched theory
from the remaining non-topological modes. We presented first
spectroscopy results with overlap fermions in quenched lattice QCD.
Finally, we compared the distribution of the smallest eigenvalue of
the Overlap-Dirac operator with the predictions from random matrix theory.

\ack{
The authors would like to thank Herbert Neuberger for
useful discussions. This research was supported by DOE contracts
DE-FG05-85ER250000 and DE-FG05-96ER40979.  Computations were performed
on the QCDSP, CM-2, and the workstation cluster at SCRI, and the Xolas
computing cluster at MIT's Laboratory for Computing Science.
}

\end{document}